\documentclass[twocolumn,showpacs,preprintnumbers,amsmath,amssymb,superscriptaddress]{revtex4}
\usepackage{graphicx}
\usepackage{dcolumn}
\usepackage{bm}
\newcommand{\bmu}{\mbox{\boldmath{$\mu$}}}
\newcommand{\blambda}{\mbox{\boldmath{$\lambda$}}}
\begin{document}
%\preprint{Draft, please do not circulate}
\title{Generalized Eulerian-Lagrangian description of Navier-Stokes and resistive MHD dynamics}
\author{Carlos Cartes}
\affiliation{Laboratoire de Physique Statistique de l'Ecole Normale Sup\'erieure, associ\'e au CNRS et
aux Universit\'es Paris VI et VII, 24 Rue Lhomond, 75231 Paris, France}
\author{Miguel D. Bustamante}
\affiliation{{Mathematics Institute, University of Warwick, Coventry CV4 7AL, United Kingdom}}
\author{Annick Pouquet}
\affiliation{National Center for Atmospheric Research, Boulder (CO) 80304 - USA}
\author{Marc E. Brachet}
\affiliation{Laboratoire de Physique Statistique de l'Ecole Normale Sup\'erieure, associ\'e au CNRS et
aux Universit\'es Paris VI et VII, 24 Rue Lhomond, 75231 Paris, France and National Center for Atmospheric Research, Boulder (CO) 80304 - USA}
\begin{abstract}
New generalized equations of motion for the Weber-Clebsch potentials that describe both the Navier-Stokes and MHD dynamics are derived. These
depend on a new parameter, which has dimensions of time for Navier-Stokes and inverse velocity for MHD. Direct numerical simulations (DNS) are
performed. For Navier-Stokes, the generalized formalism captures the intense reconnection of vortices of the Boratav, Pelz and Zabusky flow, in
agreement with the previous study by Ohkitani and Constantin. For MHD, the new formalism is used to detect magnetic reconnection in several
flows: the $3$D Arnold, Beltrami and Childress (ABC) flow and the ($2$D and $3$D) Orszag-Tang vortex. It is concluded that periods of intense
activity in the magnetic enstrophy are correlated with periods of increasingly frequent resettings. Finally, the positive correlation between the
sharpness of the increase in resetting frequency and the spatial localization of the reconnection region is discussed.

\end{abstract}
 \maketitle

\section{Introduction}
The Eulerian-Lagrangian formulation of the (inviscid) Euler dynamics in terms of advected Weber-Clebsch potentials \cite{ConstantinLocal} was extended by
Constantin \cite{Constantin} to cover the (viscous) Navier-Stokes dynamics. Ohkitani and Constantin (OC) \cite{Ohk} then performed numerical
studies of this formulation of the Navier-Stokes equations. They concluded that the diffusive Lagrangian map becomes noninvertible under time
evolution and requires resetting for its calculation. They proposed the observed sharp increase of the frequency of resettings as a new
diagnostic of vortex reconnection.

We were able to recently complement these results, using an approach that is based on a generalized set of equations of motion for the
Weber-Clebsch potentials that turned out to depend on a parameter $\tau$
which has the unit of time for the Navier-Stokes case \cite{CBB07} (the MHD case is different, see below Section \ref{sec: MPsec}).
The OC formulation is the (singular) $\tau \to 0$ limit case of our generalized formulation. Using direct numerical simulations (DNS) of the
viscous Taylor-Green vortex \cite{TG1937} we found that for $\tau\ne0$ the Navier-Stokes dynamics was well reproduced at small enough Reynolds
numbers {\it without} resetting. However, performing resettings allowed computation at much higher Reynolds number.

The aim of the present article is to extend these results to different flows,
both in the Navier-Stokes case and in magnetohydrodynamics,
and thereby obtain a new diagnostic for {\it magnetic} reconnection.
Our main conclusion is that intense reconnection of magnetic field lines is indeed captured in our new generalized formulation as a sharp
increase of the frequency of resettings. Here follows a summary of our principal results.

We first derive new generalized equations of motion for the Weber-Clebsch potentials that describe both the Navier-Stokes and MHD dynamics.
Performing DNS of the Boratav, Pelz and Zabusky flow \cite{Peltz1992}, that was previously used by Ohkitani and Constantin \cite{Ohk}, we first
check that our generalized formalism captures the intense Navier-Stokes vortex reconnection of this flow. We demonstrate the reconnection of
vortices is actually occurring at the instant of intense activity in the enstrophy, near the lows of the determinant that trigger the resettings.
We then study the correlation of magnetic reconnection with increase of resetting frequency by performing DNS of several prototypical MHD flows:
the $3$D Arnold, Beltrami and Childress (ABC) flow \cite{Archontis} and the Orszag-Tang vortex in $2$D \cite{OT2D} and $3$D \cite{OT3D}.

\section{Theoretical Framework}\label{Theory}

\subsection{General Setting}

\subsubsection{Weber-Clebsch representation for a class of evolution equations}

Let us consider a $3$D vector field $\mathbf{Z}$ depending on time and ($3$-dimensional) space, with coordinates $(x^1,x^2,x^3,t)$. Assume
$\mathbf{Z}$ satisfies an evolution equation of the kind:
\begin{eqnarray}
\label{eq:Z} \frac{D \mathbf{Z}}{D t}&=&- \nabla P + \sum_{\alpha=1}^3 u_\alpha \nabla {Z}_\alpha + \kappa \triangle \mathbf{Z}\,\\
\label{eq: divergence}\nabla \cdot \mathbf{Z} &=& 0\,,
\end{eqnarray}
where greek indices $\alpha, \beta$ denote vector field components running from $1$ to $3$, $\mathbf{u}$ is a given $3$D velocity field and we
have used the convective derivative defined by
\begin{equation}\nonumber
 \frac{D}{D t} \equiv \frac{\partial}{\partial t}+ (\mathbf{u}\cdot
 \nabla)\,.
\end{equation}

In the following sections, two different cases will be considered. In section \ref{sec: NS case} (Navier-Stokes case) the vector field
$\mathbf{Z}$ will correspond to the velocity field $\mathbf{u}$, whereas in section \ref{sec: MHD case}  (MHD case) it will correspond to the
magnetic vector potential $\mathbf{A}$.

Let us first recall that performing a change from Lagrangian to Eulerian
coordinates on the Weber transformation  \cite{Lamb}
leads to a description of the Euler equations as a system of three
coupled active vector equations in a form
that generalizes the Clebsch variable representation \cite{ConstantinLocal}.

Our starting point will be to apply this classical Weber-Clebsch representation
to the field $\mathbf{Z}$:

\begin{eqnarray} \label{eq:WC_Z} \mathbf{Z} &=& \sum_{i=1}^3 \lambda^i
\nabla \mu^i - \nabla \phi\,,
\end{eqnarray}
where each element of the $3$ pairs of Weber-Clebsch potentials $(\lambda^i, \mu^i), \,\, i=1,2,3$ is a scalar function.

Performing a variation on the Weber-Clebsch representation (\ref{eq:WC_Z}) yields the relation
\begin{equation}\label{eq: delta(Clebsch transformation)}
\mathbf{\delta Z} = \sum_{i=1}^3 \left(\delta \lambda^i \nabla \mu^i - \delta \mu^i \nabla \lambda^i\right) - \nabla \left(\delta \phi-
\sum_{i=1}^3 \delta \mu^i \lambda^i \right)\,,
\end{equation}
where the symbol $\delta$ stands for any (spatial or temporal) partial derivative.
Taking into account the identity \mbox{$\left[\nabla, \frac{D}{D t} \right] \equiv \left(\nabla \mathbf{u} \right) \cdot \nabla  $}, it is
straightforward to derive from (\ref{eq: delta(Clebsch transformation)}) the following explicit expression for the convective derivative of the
vector field $\mathbf{Z}$:
\begin{widetext}
\begin{equation}\label{eq: Dt(Clebsch transformation)}
\frac{D \mathbf{Z}}{D t} = \sum_{i=1}^3 \left(\frac{D \lambda^i}{D t} \nabla \mu^i - \frac{D \mu^i}{D t} \nabla \lambda^i\right) -
\sum_{\alpha=1}^3 Z_\alpha \nabla u_\alpha - \nabla \left(\frac{D \phi}{D t} - \sum_{i=1}^3 \frac{D \mu^i}{D t} \lambda^i \right)\,,
\end{equation}
\end{widetext}

\subsubsection{Equations of motion for the potentials}

Following steps that are similar to those presented in our previous paper \cite{CBB07}, we now derive a system of equations of motion for the
Weber-Clebsch potentials (\ref{eq:WC_Z}) that is equivalent to the original equation (\ref{eq:Z}). If we use the RHS of equation (\ref{eq:Z}) to
replace the LHS of our general identity (\ref{eq: Dt(Clebsch transformation)}), the resulting relation can be solved for the time derivative of
the potentials:
\begin{eqnarray}\label{eq: evolution_tilde Clebsch_lambda}
\frac{D \lambda^i}{D t} &=& \kappa \triangle \lambda^i +  \widetilde{L}^i[\lambda,\mu]\\
\label{eq: evolution_tilde Clebsch_mu} \frac{D\mu^i}{D t} &=&\kappa \triangle \mu^i + \widetilde{M}^i[\lambda,\mu]\,.
\end{eqnarray}
Here $\widetilde{L}^i, \widetilde{M}^i$ obey the linear equation
\begin{equation}\label{eq: linear system tilde L M}
\sum_{i=1}^3 \left(\widetilde{L}^i \nabla \mu^i - \widetilde{M}^i \nabla \lambda^i\right) =  \widetilde{\mathbf{f}}- \nabla \widetilde{G}\,,
\end{equation}
where
\begin{equation}\label{eq: tilde f}
\widetilde{\mathbf{f}} = 2 \kappa \sum_{i=1}^3 \sum_{\alpha=1}^3
\partial_\alpha \lambda^i \partial_\alpha \nabla
\mu^i
\end{equation}
and $\widetilde{G}[\lambda,\mu]$ is an arbitrary scalar related to the non-unique separation of a gradient part in eq.(\ref{eq: Dt(Clebsch
transformation)}):
\begin{equation}
\label{eq: evolution_phi} \frac{D \phi}{D t} - P = \sum_{i=1}^3\lambda^i \widetilde{M}^i - \widetilde{G} - \mathbf{u} \cdot \mathbf{Z} .
\end{equation}

The ``divergence-less gauge'' (\ref{eq: divergence}) allows one to express $\phi$ in terms of $\lambda^i$ and
$\mu^i$, as the solution of the linear equation
\begin{equation}\label{eq:lap_phi}
\triangle \phi = \sum_{i=1}^3  \nabla \cdot (\lambda^i \nabla\mu^i).
\end{equation}
Thus there is no need to solve equation (\ref{eq: evolution_phi}) for the field $\phi$, since this equation is identically
satisfied when $\phi$ is determined by eq. (\ref{eq:lap_phi}).

Equation (\ref{eq: linear system tilde L M}) above is a system of $3$ linear equations for the $6$ unknowns $\widetilde{L}^i, \widetilde{M}^i$.
When $\kappa=0$ there is a simple solution to (\ref{eq: linear system tilde L M}): $\widetilde{L}^i = \widetilde{M}^i = \widetilde{G} =0$. In
this case the evolution equations (\ref{eq: evolution_tilde Clebsch_lambda}) and (\ref{eq: evolution_tilde Clebsch_mu}) represent simple
advection.

\subsubsection{Moore-Penrose solution and minimum norm} \label{sec: MPsec}

The linear system (\ref{eq: linear system tilde L M}) is underdetermined ($3$ equations for $6$ unknowns). In order to find a solution to the
system we need to impose extra conditions. Since $\widetilde{L}^i, \widetilde{M}^i$ appear in the equations on an equal footing, it is natural to
supplement the system by a requirement of minimum norm, namely that
\begin{equation}\label{eq: vnorm}
\sum_{i=1}^3 (\widetilde{L}^i\widetilde{L}^i+{\tau^{-2}}\widetilde{M}^i\widetilde{M}^i)
\end{equation}
be the smallest possible (this is the so-called general Moore-Penrose approach \cite{Moore,Penrose,Ben-Israel}, see also our previous paper
\cite{CBB07}). The parameter $\tau$ has physical units equal to $[\widetilde{M}/\widetilde{L}]$. Using eqs.(\ref{eq: evolution_tilde
Clebsch_lambda}),(\ref{eq: evolution_tilde Clebsch_mu}) these are the units of $[\mu/\lambda]$. It will turn out (see equation (\ref{eq: mu def})
below) that $[\mu]= L$ (length) and this implies from eq.(\ref{eq:WC_Z}) that $[\lambda]=[\mathbf{Z}]$. Therefore the units of $\tau$ are
\begin{equation}
\nonumber [\tau] = \frac{L}{[\mathbf{Z}]}.
\end{equation}
In the Navier-Stokes case (section \ref{sec: NS case}) $[\mathbf{Z}]=[\mathbf{u}] = L T^{-1}$ and thus $[\tau]=T$, whereas in the MHD case (section
\ref{sec: MHD case}) $[\mathbf{Z}] = [\mathbf{A}] = L^2T^{-1}$ and thus $[\tau]=T L^{-1}$.

The Moore-Penrose solution to (\ref{eq: linear system tilde L M}),
that minimizes the norm (\ref{eq: vnorm}),
is explicitly given in equations (A6,A7) of reference \cite{CBB07}. Inserting this solution in (\ref{eq: evolution_tilde
Clebsch_lambda}),(\ref{eq: evolution_tilde Clebsch_mu}) we finally obtain the explicit evolution equations
\begin{eqnarray}\label{eq: New_Alg L}
\frac{D \lambda^i}{D t} &=& \kappa \triangle \lambda^i + \,\,\,\,\nabla \mu^i \cdot \mathbb{H}^{-1}\cdot \left(\widetilde{\mathbf{f}} - \nabla \widetilde{G}\right)\\
\label{eq: New_Alg M} \frac{D\mu^i}{D t} &=&\kappa \triangle \mu^i - \tau^{2}\,\nabla \lambda^i \cdot \mathbb{H}^{-1}\cdot
\left(\widetilde{\mathbf{f}} - \nabla \widetilde{G}\right)\,,
\end{eqnarray}
where $\widetilde{\mathbf{f}}$ is given in eq.(\ref{eq: tilde f}), the dot product denotes matrix or vector multiplication of $3$-dimensional
tensors, and $\mathbb{H}^{-1}$ is the inverse of the square symmetric $3\times 3$ matrix $\mathbb{H}$, defined by its components:
\begin{equation}\label{eq: mat}
\mathbb{H}_{\alpha \beta} \equiv \sum_{i=1}^3 \left(\tau^{2}\,\partial_\alpha \lambda^i
\partial_\beta \lambda^i+\partial_\alpha
\mu^i
\partial_\beta \mu^i\right)\,.
\end{equation}
These evolution equations together with the particular choice for the arbitrary function $\widetilde{G}$ (see equation (A11) of reference
\cite{CBB07})
\begin{equation}\label{eq: explicit G}
\widetilde{G}= \triangle^{-1}\nabla\cdot \widetilde{\mathbf{f}},
\end{equation}
is our new algorithm.

In the Navier-Stokes case, we showed in a previous paper \cite{CBB07} that the limit $\tau \to 0$ corresponds to the approach used by Ohkitani and Constantin \cite{Ohk}. In the general case (Navier-Stokes as well as MHD), we remark that the matrix $\mathbb{H}$ (see equation
(\ref{eq: mat}))
 can be written (using obvious notation) as
 $\mathbb{H} = (\nabla {\bmu})\cdot (\nabla {\bmu})^{\rm T}+\tau^2 (\nabla {\blambda})\cdot
(\nabla {\blambda})^{\rm T}$,
 which has a very simple structure in the limit
 $\tau\to 0$ .
 Because the condition $\det(\nabla {\bmu})=0$ is generically obtained at lower codimension than the
condition $\det \mathbb{H} = 0$, the limit $\tau\to 0$ is {\it singular}.

\subsection{Navier-Stokes equations}
\label{sec: NS case} The standard incompressible NS equations can be written in the form:
\begin{eqnarray}\nonumber
    \frac{D \mathbf{u}}{D t}&=& -\nabla \left(p + \frac{1}{2}|\mathbf{u}|^2\right) + \sum_{\alpha=1}^3 u_\alpha \nabla u_\alpha + \nu \triangle \mathbf{u} \\
\nonumber \nabla \cdot \mathbf{u} &=& 0\,,
\end{eqnarray}
which is indeed of the general form (\ref{eq:Z}), (\ref{eq: divergence}) with $ \mathbf{Z} = \mathbf{u}$, $\kappa = \nu$ and $P=p +
\frac{1}{2}|\mathbf{u}|^2$.

\subsection{MHD equations}
\label{sec: MHD case} The standard incompressible MHD equations for the fluid velocity $\mathbf{u}$ and the induction field $\mathbf{b}$, expressed in Alfvenic
velocity units, can be written in the form:
\begin{eqnarray}\label{eq:momentum}
    \frac{D \mathbf{u}}{D t}&=& -\nabla p + \nu \triangle \mathbf{u} + (\nabla \times \mathbf{b}) \times \mathbf{b}\\
\label{eq:B}
    \frac{D \mathbf{b}}{D t} &=& (\mathbf{b}\cdot
 \nabla)\mathbf{u} + \eta \triangle \mathbf{b} \\
\label{eq: vinc} \nabla \cdot \mathbf{u} &=& 0\\
\nonumber \nabla \cdot \mathbf{b} &=& 0\,,
\end{eqnarray}
where $\nu$ and $\eta$ are the viscosity and magnetic resistivity, respectively.

We introduce the vector potential in the Coulomb gauge:
\begin{eqnarray}
\nonumber \mathbf{b} &=& \nabla \times \mathbf{A}\\
\nonumber \nabla \cdot \mathbf{A} &=& 0\,.
\end{eqnarray}
Using the identity $\nabla \times (\sum_{\alpha=1}^3 u_\alpha \nabla {A}_\alpha - (\mathbf{u} \cdot \nabla) \mathbf{A}) = (\mathbf{b} \cdot
\nabla) \mathbf{u} - (\mathbf{u} \cdot \nabla) \mathbf{b} - (\nabla \cdot \mathbf{u}) \mathbf{b}\,$ and the incompressibility condition (\ref{eq:
vinc}), eq. (\ref{eq:B}) can be written as
\begin{eqnarray}
\nonumber
    \frac{D \mathbf{A}}{D t}&=& - \nabla c + \sum_{\alpha=1}^3 u_\alpha \nabla {A}_\alpha + \eta \triangle \mathbf{A},
\end{eqnarray}
which is indeed of the general form (\ref{eq:Z}) with $ \mathbf{Z} = \mathbf{A}$, $\kappa = \eta$ and $P = c$.

\section{Numerical Results}\label{Direct Numerical Simulations}
\subsection{Implementation}

\subsubsection{Initial conditions in pseudo-spectral method}

Spatially periodic fields can be generated from the Weber-Clebsch representation (\ref{eq:WC_Z}) by setting
\begin{equation}\label{eq: mu def}
\mu^i =  x^i + \mu^i_{\rm p},
\end{equation}
and assuming that $\mu^i_{\rm p}$ and the other fields $\lambda^i$ and
$\phi$ appearing in (\ref{eq:WC_Z})
are periodic.
Indeed, any given periodic field $\mathbf{Z}$ can
be represented in this way by setting
\begin{eqnarray}\label{eq: resetting mu}
\mu^i_{\rm p} &=&0\\
\label{eq: resetting lambda}
\lambda^i&=&Z^i\\
\label{eq: resetting phi}
\phi &=&0.
\end{eqnarray}
Note that the time independent non-periodic
part of $\mu^i$ of the form given in (\ref{eq: mu def})
is such that the {\it gradients} of $\mu^i$ are periodic.
It is easy to check that this representation
is consistent with the generalized equations of motions
(\ref{eq: New_Alg L},\ref{eq: New_Alg M}).
We chose to use standard Fourier
pseudo-spectral methods, both for their precision
and for their ease of implementation \cite{Got-Ors}.

\subsubsection{Resettings and reconnection}

Following Ohkitani and Constantin \cite{Ohk}, we now define resettings. Equations (\ref{eq: resetting mu}), (\ref{eq: resetting lambda}) and
(\ref{eq: resetting phi}) are used not only to initialize the Weber-Clebsch potentials at the start of the calculation but also to {\it reset}
them to the current value of the field $ \mathbf{Z}$, obtained from (\ref{eq:WC_Z}) and (\ref{eq:lap_phi}), whenever the minimum of the
determinant of the matrix (\ref{eq: mat}) falls below a given threshold
\begin{equation}\nonumber
\det \mathbb{H} \le\epsilon^2 .
\end{equation}

It is possible to capture reconnection events using resettings. The rationale for this approach is that reconnection events are associated to
localized, intense and increasingly fast activity which will drive the potentials to a (unphysical) singularity in a finite time. One way to
detect this singularity is via the alignment of the gradients of the potentials, which leads to the vanishing of $\det \mathbb{H}$ at the
point(s) where this intense activity or `anomalous diffusion' is taking place. Now, the time scale of this singularity is much smaller than the
time scale of the reconnection process itself \cite{Ohk}, so when $\det \mathbb{H}$ goes below the given threshold and a resetting of the
potentials is performed, the anomalous diffusion starts taking place again, more intensely as we approach the fastest reconnection period,
driving the new (reset) potentials to a new finite-time singularity, in a time scale that decreases as we approach this period. Therefore,
successive resettings will be more and more frequent near the period of fastest reconnection, and that is what we observe in the numerical
simulations. This procedure will be used to capture reconnection events in particular flows in both the Navier-Stokes case ($\mathbf{Z} =
\mathbf{u}$, Section \ref{sec:NS}) and the MHD case ($\mathbf{Z} = \mathbf{A}$, Sections \ref{sec:2DOT} and \ref{sec:3DOT}).

\subsection{Navier-Stokes case: BPZ Flow, resettings and reconnection}
\label{sec:NS}

Ohkitani and Constantin (OC) \cite{Ohk} used a flow that initially consists of two orthogonally placed vortex tubes that was previously introduced in
Boratav, Pelz and Zabusky (BPZ) \cite{Peltz1992} to study in detail vortex reconnection. Our previous numerical study of the generalized
Weber-Clebsch description of Navier-Stokes dynamics \cite{CBB07} was performed using the Taylor-Green vortex, a flow in which vorticity layers
are formed in the early stage, followed by their rolling-up by Kelvin-Helmholtz instability \cite{Brachet1}. It can be argued \cite{Ohk} that
cut-and-connect type reconnections are much more pronounced in the BPZ flow than in the Taylor-Green flow. In this section we present
comparisons, performed on the BPZ flow, of our $\tau \ne 0$ generalized algorithm with direct Navier-Stokes simulations and with OC original
approach. The potentials are integrated with resettings in resolution $128^3$ for a Reynolds number of $R=1044$, which is the one used by BPZ and
OC.

The BPZ initial data is explicitly given in \cite{Peltz1992}.

\subsubsection{Comparison of Weber-Clebsch algorithm with DNS of Navier-Stokes}

In order to characterize the precision of the $\tau \ne 0$ Weber-Clebsch algorithm, we now compare the velocity field $\mathbf{Z} = {\bf u}$
obtained from (\ref{eq:WC_Z}) and (\ref{eq:lap_phi}), by evolving the Weber-Clebsch potentials using (\ref{eq: New_Alg L})--(\ref{eq: explicit
G}), with the velocity field obtained independently by direct Navier-Stokes evolution from the BPZ initial data.

More precisely, we compare the associated kinetic enstrophy $\Omega(t)=\sum_{k}k^2 E(k,t)$ where the kinetic energy spectrum $E(k,t)$ is defined
by averaging the Fourier transform ${\bf \hat u}({\bf k',t})$ of the velocity field (\ref{eq:WC_Z}) on spherical shells of width $\Delta k = 1$,
\begin{equation}\nonumber
E(k,t) = {\frac1 2} \sum_{k-\Delta k/2< |{\bf k'}| <  k + \Delta k/2} |{\bf \hat u}({\bf k',t})|^2 \, .
\end{equation}

\begin{figure}[ht!]
\vspace{2.0cm}
\begin{center}
 \hspace{0.cm}
\includegraphics[width=8.0cm,angle=0]{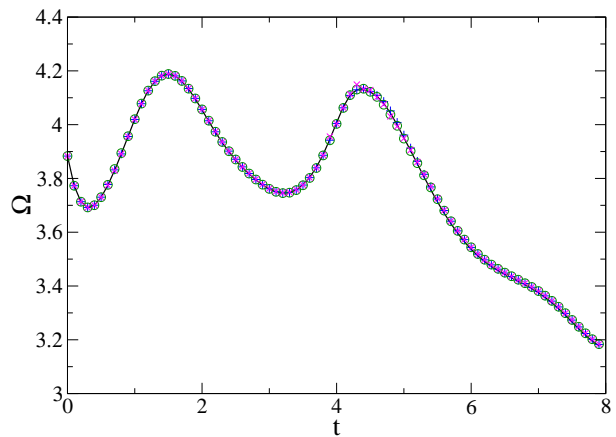}
\vspace{1.cm} \caption{\small{Navier-Stokes case: BPZ Flow. Temporal evolution
    of kinetic enstrophy $\Omega$ for a Reynolds number of $R=1044$ with $\tau= 0$, $0.01$ and $0.1$ ($+$, $\circ$ and
    $\times$). The solid line comes from a direct numerical simulation (DNS) at resolution $128^3$. } \label{EnstrophyBPZ}}
\end{center}
\end{figure}

\begin{figure}[ht!]
\vspace{0.5cm}
\begin{center}
 \hspace{0.cm}
\includegraphics[width=8.0cm,angle=0]{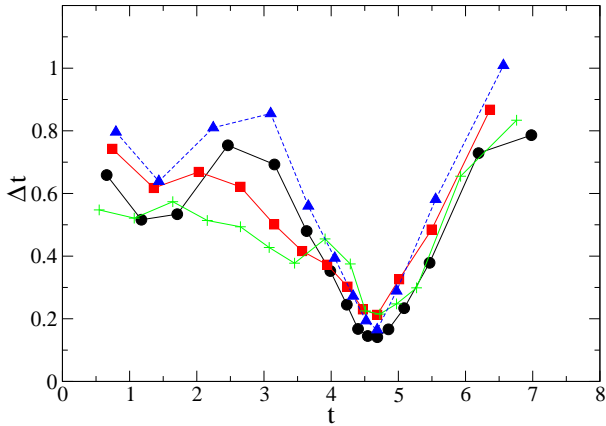}
\vspace{0.cm} \caption{\small{Navier-Stokes case: BPZ flow. Temporal evolution of resetting interval $\Delta t$ for $\tau= 0$, $0.01$ and $0.1$  ($\circ$, $\square$
and $+$), the triangles correspond to the simulation performed by Ohkitani and Constantin. } \label{DeltaResBPZ}}
\end{center}
\end{figure}

Figure \ref{EnstrophyBPZ} shows that the kinetic enstrophy is well resolved, independently of the choice of the parameter $\tau$.

\subsubsection{Time between resettings as a method for reconnection capture}

In this section we study the influence of the parameter $\tau$ on the temporal distribution of the intervals $\Delta t_j = t_j - t_{j-1}$ between
resetting times $t_j$, at fixed value of the resetting threshold $\epsilon^2 = 0.1$. Using the same Reynolds number and resolution that was used
to create Fig. \ref{EnstrophyBPZ}, Figure \ref{DeltaResBPZ} is a plot of $\Delta t$ as a function of time, for simulations with different values
of $\tau$. In the same figure we also show the corresponding $\Delta t$ for a replica of the simulation performed by OC, that is in excellent
agreement with our general case.

\begin{figure}[ht!]
\vspace{0.0cm}
\begin{center}
 \hspace{0.cm}
\includegraphics[width=8.0cm,angle=0]{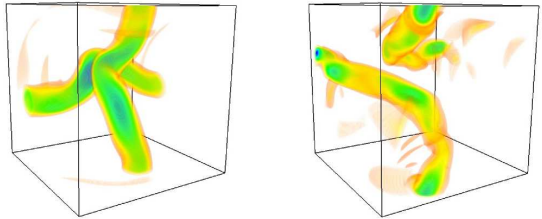}
\vspace{0.cm} \caption{\small{Visualization of vorticity $\omega$ for the BPZ flow (same conditions than in Figures \ref{EnstrophyBPZ} and
\ref{DeltaResBPZ}). Note the change of topology of the vortex tubes before
(left, $t=3.2$, $\omega_\textrm{max}=20$) and after (right, $t=7.1$,
$\omega_\textrm{max}=15$) the reconnection process. Isosurfaces colors: orange: $6$,
yellow: $9$, green: $12$ and blue: $16$ (color online). }
\label{FiguraRecon}}
\end{center}
\end{figure}

We see that, independently of $\tau$, there are sharp minima in $\Delta t$ during the periods of maximum enstrophy (see Fig. \ref{EnstrophyBPZ}).
Inspection of figure \ref{FiguraRecon} demonstrates that  the deepest minimum corresponds in fact to the time when reconnection is taking place.
The main tubes in the left and right figures are isosurfaces of vorticity corresponding to $60\%$ of the maximum vorticity, which is attained
inside each of the main tubes.

\begin{figure}[ht!]
\vspace{0.0cm}
\begin{center}
 \hspace{0.cm}
\includegraphics[width=8.0cm,angle=0]{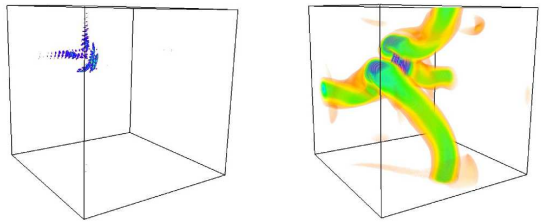}
\vspace{0.cm} \caption{\small{Visualization of the determinant of the matrix (\ref{eq: mat}) (left)
and vorticity (right, with the same color map than in Figure \ref{FiguraRecon}) at reconnection time
$t=4.7$, $\omega_\textrm{max}=43.6$ (see Figure \ref{DeltaResBPZ}), for $\tau=0.01$.
The region where the determinant triggers resetting is within the
displayed blue isosurface at $9$ times the triggering level $\epsilon^2$
(color online).} \label{FiguraDet}}
\end{center}
\end{figure}

Figure \ref{FiguraDet} (left) shows that the spatial region where the determinant $\det \mathbb{H}$ goes below the threshold before each
resetting corresponds to a small, localized neighborhood between the main interacting vortices. This region is seen in the right figure as a
bridge connecting the two vortices: this bridge is an isosurface of vorticity corresponding to $73\%$ of the maximum vorticity, which is attained
inside the bridge. The main tubes correspond to isosurfaces of $30\%$ of the maximum vorticity. Note that this behavior of the determinant $\det
\mathbb{H}$ is also true for any value of $\tau$ (data not shown), confirming in this way the original rationale for the study of reconnection
with the aid of resettings.

Figures \ref{FiguraRecon} and \ref{FiguraDet} were made using
the VAPOR \cite{ncarVapor1,ncarVapor2}visualization software.

\subsection{MHD Flows}

In this section we study MHD flows with simple initial conditions. The magnetic potential $\mathbf{Z} = \mathbf{A}$ is obtained in terms of the
Weber-Clebsch potentials from (\ref{eq:WC_Z}) and (\ref{eq:lap_phi}), and the Weber-Clebsch potentials are evolved using equations (\ref{eq:
New_Alg L})--(\ref{eq: explicit G}).

We treat the evolution of the velocity field in two different ways: (i) As a kinematic dynamo (ABC flow, Section \ref{sec:ABC}), where the
velocity is kept constant in time; (ii) Using the full MHD equations (Orszag-Tang $2$D and $3$D, Sections \ref{sec:2DOT} and \ref{sec:3DOT}),
where the velocity field is evolved using the momentum equation (\ref{eq:momentum}).

To compare with DNS of the induction equation (\ref{eq:B}) for the magnetic field we proceed analogously as in the Navier-Stokes case. We compare
the magnetic enstrophy \cite{Dahl89} $\Omega_m(t) = \sum_{k}k^2 E_m(k,t)$, where the magnetic energy spectrum $E_m(k,t)$ is defined by averaging
the Fourier transform ${\bf \hat b}({\bf k',t})$ of the magnetic field $\mathbf{b} = \nabla \times \mathbf{A}$ (with $\mathbf{A}$ given by
(\ref{eq:WC_Z})) on spherical shells of width $\Delta k = 1$,
\begin{equation}\nonumber
E_m(k,t) = {\frac1 2} \sum_{k-\Delta k/2< |{\bf k'}| <  k + \Delta k/2} |{\bf \hat b}({\bf k',t})|^2 \, .
\end{equation}
Note that magnetic dissipation is the square current.

Resettings will be performed with a resetting threshold $\epsilon^2 = 0.1$. We have checked that $\epsilon^2=0.4$ and $\epsilon^2=0.025$ give results that
vary only slightly (figures not shown). This is an evidence of the robustness of the resetting method and a validation of the rationale for the
use of resettings to diagnose reconnection.

\subsubsection{Kinematic dynamo: ABC Flow}
\label{sec:ABC}
We have used the ABC \cite{Archontis} velocity:
\begin{eqnarray}
u_x&=& B_0 \cos k_0 y + C_0 \sin k_0 z  \nonumber\\
u_y&=& C_0 \cos k_0 z + A_0 \sin k_0 x  \nonumber\\
u_z &=& A_0 \cos k_0 x + B_0 \sin k_0 y  \,, \nonumber
\end{eqnarray}
with $k_0=2$ and $A_0=B_0=C_0=1$. We used an initial magnetic seed that reads
\begin{eqnarray}
A_x&=&0 \nonumber\\
A_y&=& 0 \nonumber\\
A_z &=& d_0 \sin x \sin y \nonumber\,.
\end{eqnarray}
The magnetic resistivity has been chosen as $\eta = 1/12$ and we have set $d_0 = 1/100$ for simplicity (its value is unimportant in the kinematic
dynamo).

\begin{figure}[ht!]
\vspace{0.cm}
\begin{center}
 \hspace{0.cm}
\includegraphics[width=8.0cm,angle=0]{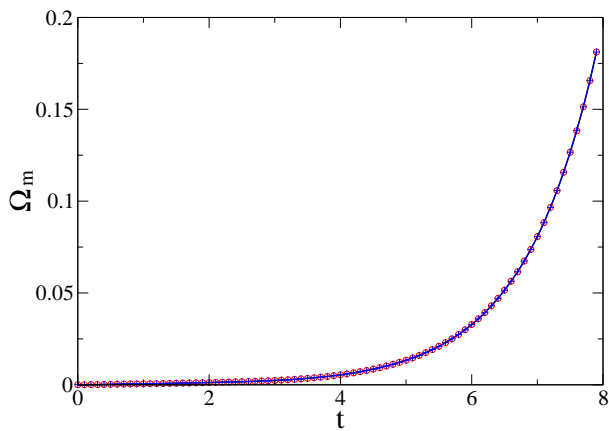}
\vspace{.cm} \caption{\small{Temporal evolution of magnetic enstrophy $\Omega_m$ for the ABC flow (kinematic dynamo) with $\tau=0$ and $1$
($\circ$ and $+$), with a resolution of $128^3$ and the constants $\nu=0$ and $\eta=1/12$. The solid line comes from a direct numerical
simulation (DNS) of the induction equation for the magnetic field. } \label{fig_Magnetic_Enstr_ABC}}
\end{center}
\end{figure}

Runs with resettings are compared for different values of the parameter $\tau$. It is seen in Fig. \ref{fig_Magnetic_Enstr_ABC} that the magnetic
enstrophy $\Omega_m$ is well resolved for each case, at resolution $128^3$.

The resettings are quite regular in time and indeed they slow down as time goes by, at a regular rate which decreases with increasing resolution
(figure not shown). There is no increase in the resetting frequency. This behavior is consistent with the monotonic behavior of the magnetic
enstrophy and with the absence of localized or intense activity of the magnetic field.

\subsubsection{Full MHD equations: 2D Orszag-Tang Vortex}
\label{sec:2DOT}

In the rest of the paper, the full MHD equations of motion are integrated. The momentum equation for the velocity (\ref{eq:momentum}) is
integrated together with the Weber Clebsch evolution equations (\ref{eq: New_Alg L})--(\ref{eq: explicit G}) where the magnetic potential
$\mathbf{Z} = \mathbf{A}$ is obtained from (\ref{eq:WC_Z}) and (\ref{eq:lap_phi}).

We have chosen the following initial data for the $2$D Orszag-Tang
(hereafter, OT)
vortex \cite{OT2D}:
\begin{eqnarray}
u_x&=& -2 \sin y \nonumber\\
u_y&=&  2 \sin x \nonumber\\
u_z &=& 0, \nonumber\\
A_x&=& 0 \nonumber\\
A_y&=& 0 \nonumber\\
A_z &=& 2\cos x \cos 2y \,.  \nonumber
\end{eqnarray}
The OT vortex has a magnetic hyperbolic X-point located at a stagnation point of the velocity, and is a standard test of magnetic reconnection, both in two dimensions \cite{add1} and in three dimensions \cite{add2}, see below section \ref{sec:3DOT}.

\begin{figure}[ht!]
\vspace{0.cm}
\begin{center}
 \hspace{0.cm}
\includegraphics[width=8.0cm,angle=0]{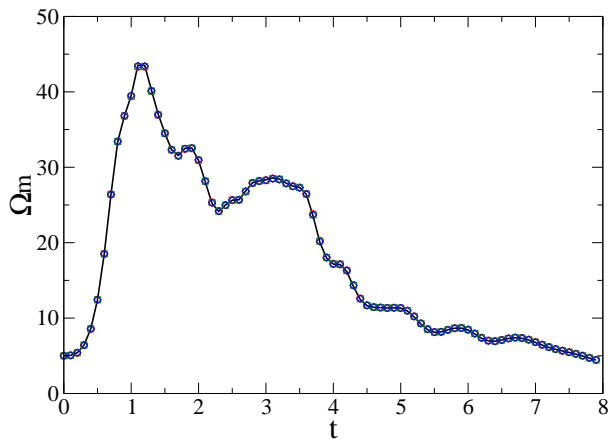}
\vspace{0.cm} \caption{\small{Temporal evolution of magnetic enstrophy $\Omega_m$ for
    Orszag-Tang in 2D for $\tau=0$, $0.01$ and $1$ ($\circ$, $\square$ and $\diamond$) with a
    resolution of $128^2$ and $\eta=\nu=0.005$. Solid line: Direct numerical simulation of MHD equations.} \label{Current2d}}
\end{center}
\end{figure}

We compare runs with resettings for different values of the parameter $\tau$. Figure \ref{Current2d} shows that the magnetic enstrophy is well
resolved in resolution $128^2$.

\begin{figure}[ht!]
\vspace{0.cm}
\begin{center}
 \hspace{0.cm}
\includegraphics[width=8.0cm,angle=0]{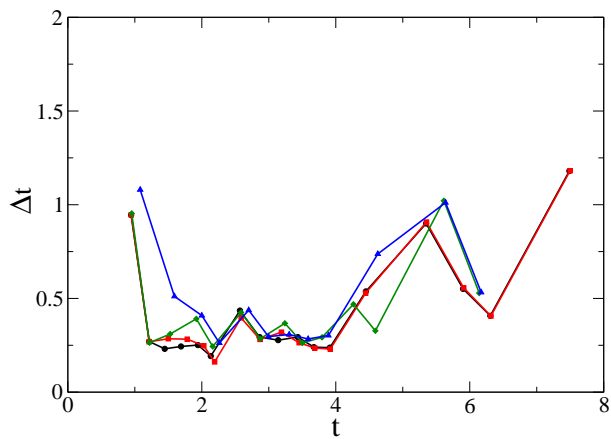}
\vspace{0.cm} \caption{\small{Temporal evolution  of $\Delta t$ for $\tau= 0$,
   $0.01$, $0.1$ and $1$
    ($\circ$, $\square$,  $\diamond$ and $\triangle$), for a simulation of Orszag-Tang in
    2D with $\nu=\eta=0.005$ and a resolution of $128^2$. } \label{DeltaResOT2d}}
\end{center}
\end{figure}

Figure \ref{DeltaResOT2d} shows the time between resettings as a function of time, for runs performed with different values of $\tau$. It is
apparent from the figure that there are periods of frequent resettings which coincide with the periods of high magnetic enstrophy from Fig.
\ref{Current2d}. This is a robust evidence of the utility of the resetting approach for $2$D magnetic reconnection.

We have also simulated the Orszag-Tang vortex in the so-called $2.5$D setting \cite{Montgomery82} (see also the DiPerna-Majda's construction
\cite{DiPerna87}), defined by the same initial data as the above $2$D Orszag-Tang vortex, but with $A_x= \sin y $ and $A_y=- \sin x$. We obtained
(data not shown) a behavior of the resetting frequency which was very similar to that of the $2$D case.

\subsubsection{Full MHD equations: 3D Orszag-Tang Vortex}
\label{sec:3DOT}

\begin{figure}[ht!]
\vspace{1.cm}
\begin{center}
 \hspace{0.cm}
\includegraphics[width=8.0cm,angle=0]{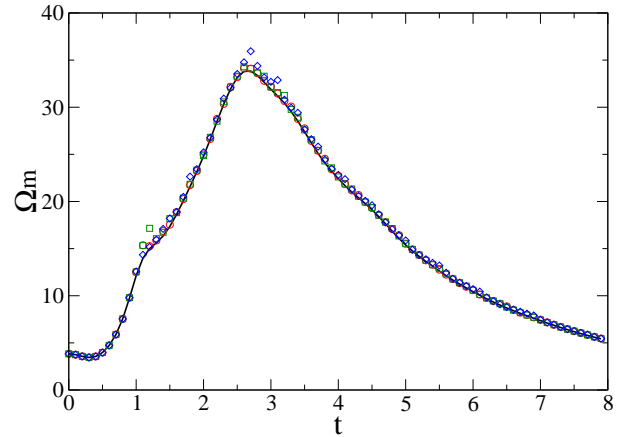}
\vspace{0.cm} \caption{\small{Temporal evolution of magnetic enstrophy $\Omega_m$ for
    Orszag-Tang in 3D for $\tau=0$, $0.1$ and $1$ ($\circ$, $\square$ and $\diamond$) with a
    resolution of $128^3$ and $\eta=\nu=0.005$. Solid line: Direct numerical simulation of MHD equations. } \label{OmegaM Orszag-Tang3d}}
\end{center}
\end{figure}

\begin{figure}[ht!]
\vspace{0.cm}
\begin{center}
 \hspace{0.cm}
\includegraphics[width=8.0cm,angle=0]{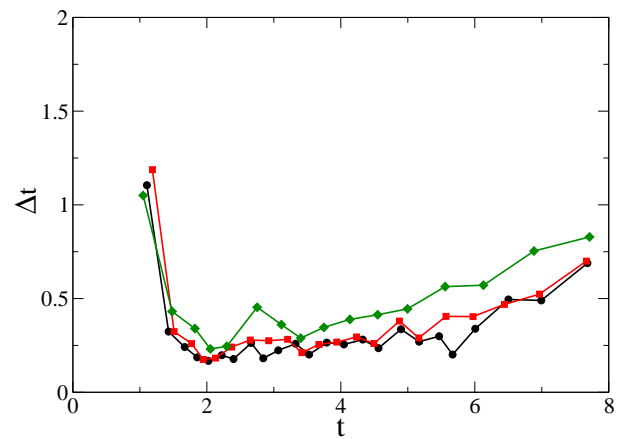}
\vspace{0.cm} \caption{\small{Temporal evolution of $\Delta t$ for $\tau= 0$, $0.1$ and $1$
    ($\circ$, $\square$ and $\diamond$), for a simulation of Orszag-Tang in
    3D. } \label{DeltaResOT3d}}\end{center}
\end{figure}

For the $3$D Orszag-Tang vortex \cite{OT3D} the initial magnetic potential reads
\begin{eqnarray}
A_x&=& c_0 \left(\cos y - \cos z \right) \nonumber\\
A_y&=& c_0 \left(-\cos x + \cos z \right) \nonumber\\
A_z &=& c_0 \left(\cos x + \cos 2y \right)\,, \nonumber
\end{eqnarray}
with $c_0=0.8$. The initial velocity is given by
\begin{eqnarray}
u_x&=& -\sin y  \nonumber\\
u_y&=& \sin x  \nonumber\\
u_z &=& 0 \,.\nonumber
\end{eqnarray}
As in the $2$D case, we compare runs with resettings for different values of the parameter $\tau$. Figure \ref{OmegaM Orszag-Tang3d} shows that
the magnetic enstrophy is well resolved in resolution $128^3$, and Fig. \ref{DeltaResOT3d} shows the time between resettings as a function of
time. Again the periods of frequent resettings coincide with the periods of high magnetic enstrophy from Fig. \ref{OmegaM Orszag-Tang3d}, proving
the utility of the resetting approach for $3$D magnetic reconnection.

\section{Conclusions}

We have shown that the generalized Weber-Clebsch evolution equations allow to study reconnection events for both Navier-Stokes and MHD dynamics.
We have checked for the Navier-Stokes BPZ flow that reconnection events can be viewed as periods of fast and localized changes in the geometry of
the Weber-Clebsch potentials, leading to more and more frequent resetting of the potentials.

We have applied the new generalized Weber-Clebsch evolution equations to the study of magnetic reconnection in MHD.
Taking as examples both the $2$D and
$3$D Orszag-Tang vortices, we show a correlation of the reconnection events (associated to periods of high magnetic dissipation) with the periods of
fast changes in the geometry of the Weber-Clebsch potentials, leading to frequent resettings of the potentials.

However, unlike the case of BPZ reconnection, in this case the frequency of resettings does not have a sharp peak but a smeared one. Notice that,
in the Navier-Stokes case, the corresponding frequency of resettings for the Taylor-Green vortex has also a mild peak. \cite{CBB07} One can argue
that the 2D and 3D Orszag-Tang flows are more similar to Taylor-Green than to BPZ. Indeed, both Orszag-Tang and Taylor-Green have initial
conditions with just a few Fourier modes, therefore they are extended spatially, whereas the BPZ initial condition is spatially localized (two
orthogonal vortex tubes).

This wide spatial extent of the vorticity in both Orszag-Tang and Taylor-Green vortices, as opposed to the localized extent of BPZ, might be the reason
for the mildness in the shape of the minimum of the time between resettings. In both spatially extended cases one expects reconnection events to
happen in relatively distant places at similar times, as opposed to the BPZ very localized cut-and-connect type of reconnection. In terms of the
singularities of the Weber-Clebsch potentials and associated resetting, we should observe (to be studied in detail in future work) that the set of points
where $\det \mathbb{H}$ goes below the threshold consists of an extended region, as opposed to BPZ where we have confirmed that these points
belong to a very localized region in space. Consequently, the widely distributed events that lead to resetting in Orszag-Tang and Taylor-Green configurations
would tend to be less correlated in time, leading to the smearing of the minimum of the curve for the time between resettings, which would
otherwise be very sharp if the events were more localized and therefore more correlated in time.

{\bf Acknowledgments:}
We acknowledge very useful scientific discussions with
Peter Constantin and Edriss S. Titi.
One of the authors (MEB.) acknowledges support from an ECOS/CONICYT action.
The computations were carried out at
the Institut du D\'eveloppement et des Ressources en
Informatique Scientifique (IDRIS) of the Centre National pour la Recherche
Scientifique (CNRS).

%\appendix

\bibliographystyle{unsrt}
%\bibliography{bibli}

\end{document}